\documentclass[iop,apj,tighten,twocolappendix,numberedappendix]{emulateapj}

\usepackage{graphicx}
\usepackage{amsmath}
\usepackage{amssymb}
\usepackage{color}

\usepackage[breaklinks,colorlinks,citecolor=blue,linkcolor=red]{hyperref} 
\usepackage[all]{hypcap}

\begin{document}

\title{Steeper stellar cusps in galactic centers from binary disruption}

\author{Giacomo Fragione\altaffilmark{1} and Re'em Sari\altaffilmark{1}}
\affil{$^1$Racah Institute for Physics, The Hebrew University, Jerusalem 91904, Israel}

\email{giacomo.fragione@mail.huji.ac.il\\sari@phys.huji.ac.il}
 
\begin{abstract}
The relaxed distribution of stars around a massive black hole is known to follow a cusp profile $\rho(r)\propto r^{-\alpha}$ with characteristic slope $\alpha=7/4$. This follows from energy conservation and a scattering rate as given by two body encounters. However, we show that injection of stars close to the black hole, i.e. a source term in the standard cusp picture, modifies this profile. In the steady-state configuration, the cusp develops a central region with typical slope $\alpha=9/4$ in which stars diffuse outward. Binary disruption by the intense tidal field of the massive black hole is among the phenomena that take place in the Galactic Center. In such disruption, one of the binary members remains bound to the black hole, thus providing a source term of stars close to the black hole. Assuming a binary fraction of $0.1$ and an orbital circularization efficiency of $0.35$, we show that this source is strong enough to modify the cusp profile within $\approx 0.07$ pc in the Galactic Center. If the binary fraction at the influence radius is of order unity and the orbits of all the captured stars are efficiently circularized, the steeper cusp extends almost as far as the radius of influence of the black hole.
\end{abstract}


\section{Introduction}

Super massive black holes (SMBHs), with masses in the range $10^5\ \mathrm{M}_{\odot}\lesssim M_{BH} \lesssim 10^{10}\ \mathrm{M}_{\odot}$, lie in the inner regions of most, if not all, galaxies. Such MBHs are embedded in dense and complex structures including stars in disk-like and spherical configurations and diffuse gas (see review \citet{ale17} for details). Among all the galactic nuclei, the Milky Way's Galactic Center (GC) has a singular and fundamental role in modern physics due to its relatively close distance ($\approx 8.15$ kpc). Present day technology, including the high angular resolution observations with the \textit{Hubble Space Telescope}, allows us to resolve physical scales on the order of fraction of pc and thus to study the physical properties and dynamics of individual stars \citep{sch14}. Developing a solid theoretical picture of the phenomena we observe in the GC with unprecedented telescopes gives us the opportunity of improving not only of the understanding of our Galaxy, but also of the evolution and structure of dense galactic nuclei in general.

Due to the large mass ratio between the SMBH and stars, stars can be approximated as test masses moving on keplerian orbits in the spherical smooth potential of the SMBH with typical periods $\lesssim 10^5$ yr \citep{ale17}. Such approximation holds up to nearly the so-called gravitational influence radius $r_h$, defined as the radius at which the potential of the SMBH becomes comparable to the average galactic field or the radius of a sphere that encloses a mass in stars similar to $M_{BH}$ \citep{mer13}. In the case of the Milky Way, $r_h\approx 2$ pc. Stars' potential becomes relevant on longer timescales. The cumulative effect of uncorrelated $2$-body gravitational interactions randomizes both the orbital energy and angular momentum on the $2$-body relaxation timescale \citep{bar14,bar16}
\begin{equation}
T_{2b}(a)=\frac{C_{2b}}{N(a)}\left(\frac{M_{BH}}{m}\right)^2 P(a)\ ,
\label{eqn:t2b}
\end{equation}
where $C_{2b}$ is a dimensionless constant that includes the Coulomb logarithm \citep{bin08}, $N$ is the number of particles, $m$ is the typical star mass and $P(a)$ is the Keplerian period of an object moving on an orbit of semimajor axis $a$. In the Milkw Way's GC, $T_h=T_{2b}(a=r_h)\approx 3\times 10^{10}$ yr at the influence radius.

Equation \ref{eqn:t2b} gives the typical timescale it takes for a system to evolve as a consequence of the sum of many weak, uncorrelated two-body interactions \citep{mer13}. \citet{baw76} were the first to solve the Fokker-Plack equation related to such a scenario, finding a density profile of single mass stars $n\propto r^{-7/4}$, where $r$ is the radial distance with respect to the SMBH. $T_{2b}\gtrsim 10^9$-$10^{10}$ yr and may be orders of magnitude larger than the age of the Universe, becoming negligible. However, galactic nuclei with $M_{BH}\lesssim 10^7\ \mathrm{M}_{\odot}$ have two-body timescale small enough to make the effects of the uncorrelated stellar interactions important within a Hubble time. Such collisional systems are also of interest for the loss-cone dynamics, gravitational waves and tidal dissipation \citep{hop04,hop05,hop06,hoa06,ale09,sto13}. 

Another mechanism for transporting stars in the galactic center is enabled by binaries. Binary stars in the inner regions of the GC may undergo three-body exchange interactions with the SMBH. \citet{hil88} was the first to describe such a scenario, where one of the stars is expelled from the GC with velocities of hundreds km s$^{-1}$, becoming a hypervelocity star (HVS), while the other one remains bound to the SMBH \citep{yut03,sar10,kob12,ros14}. Other mechanisms have been proposed to explain HVSs, such as star clusters-SMBH interactions \citep{cap15,fra16,fck17} and supernova explosions \citep{zub13}, but the tidal disruption of binaries is commonly accepted as the most relevant channel. HVSs can provide information about the Galactic mass distribution and potential triaxiality \citep*{gne05,frl17,ros17}, but also on planetary dynamics in extreme conditions \citep*{gin12,frg17}.

In the Hill's scenario, a lot of effort has been spent or on the origin of the dynamics of the three-body interaction and on the information we may achieve from HVS data \citep{brw15}. Apart from the claims that some of the S-stars in the GC, see e.g. \citet{gou03}, may have been originated as a consequence of binary tidal disruption, little attention has been dedicated to the long-term feedback on the stellar cusp by the former companions of HVSs. The breakup of such a binary typical create a bound star whose semimajor axis is order of magnitude smaller than the original semimajor axis of the binary around the black hole. In this paper, we address the question whether such a continuous supply of stars from the radius of influence into a range of much tighter orbits may modify the standard \citet{baw76} picture. We treat the process of binary breakup by adding a source term at the semimajor axis of the bound star. We then use theoretical arguments and numerical simulations to show that such source term changes the slope of the cusp external to the source position. 

The paper is organized as follows. In Section 2, we describe the theoretical framework, in which we use theoretical arguments to show that introducing a source term for the general \citet{baw76} makes the the density profile more cuspy from $n(r) \propto r^{-7/4}$ to $n(r) \propto r^{-9/4}$. In Section 3 we introduce the computational algorithm used to simulate our astrophysical scenario, while, in Section 4, we present our results. In Section 5, we discuss the implications of our study. Finally, in Section 6, we draw our conclusions.

\section{Theoretical framework}

\citet{baw76} derived the integro-differential equation describing the dynamical evolution of stars inside the influence sphere of a SMBH. In the steady state case, the equation allows a simple power-law solution for the density of stars
\begin{equation}
n(r)=A r^{-7/4}\ ,
\end{equation}
where $r$ is the radial distance from the central SMBH. Such a solution can be understood in terms of the energy flux and of the typical timescale on which such a flux is propagated through the stellar cusp. The star typical energy is $E_*=-GM_{BH}m/2r$. As stars scatter each other in semimajor axis, a loss of negative energy needs a flow of positive energy out through the cusp in a steady state. The typical timescale of energy flow is the $2$-body timescale (Eq. \ref{eqn:t2b}), during which the $N(r)$ cusp stars can carry an energy of order $N(r)E(r)$ through the shell at radius $r$ \citep{baw76,bin08}. Since the energy flow must be independent on the radius in a steady state, and assuming a power-law distribution for stars $N(r)\propto \rho(r) r^3 \propto r^{-\alpha+3}$, 
\begin{equation}
\mathcal{F}=\frac{N(r)E(r)}{T_{2b}(r)}\propto r^{-2\alpha+7/2}=\mathrm{const}\ .
\label{eqn:cuspnosour}
\end{equation}
The above equation gives the standard result $\alpha=7/4$ for the distribution of stars around a massive object \citep{baw76}.

\subsection{Steady State With a Source Term}
In the case in which we have a source term, we introduce stars with a rate $\dot{N}$. Let $r_{source}$ be the distance at which we inject these stars. The stars will scatter each other via $2$-body process, transporting themselves away from the injection point. If the rate of injection is significant enough,
we expect that the flux would be in both inward and outward directions. The inward flux is identical to the standard case, and, as explained above, a constant energy flow dictates $\rho \propto r^{-7/4}$ for $r<r_{source}$. However, for the outward flow of stars from the injection point $r>r_{source}$, energy conservation is less restrictive than stellar number conservation. The energy flux would be very small, with a constant stellar flux given by
\begin{equation}
\dot{N}_{inj}=\frac{N(r)}{T_{2b}(r)}\propto r^{-2\alpha+9/2}=const.\ .
\label{eqn:cuspsour}
\end{equation}
Therefore, $\alpha=9/4$, in the outer cusp. Note, that a slope of $\alpha=9/4$ was first derived by \citet{pee72}, but it was then concluded to be erroneous by \citet{sha76} and \citet{baw76}. For elastic scattering between particles, such a slope results in outward rather than inward flux.  More recently, \citet{sag06} arrived at the same density profile $\alpha=9/4$ for inward flow in case of non elastic physical collisions.

\subsection{Local v.s. Non Local Interactions}

In the derivation of the steady state solutions above, both with and without a source term, we focused on ``local" interactions, i.e. interactions between stars of similar semimajor axis. However, stars with highly eccentric orbits have the opportunity to interact with stars on smaller semimajor axis. Since the density of stars closer to the SMBH is larger, these interactions may have some importance. The cross section for an interaction that changes the energy by $\Delta E$ is given by $\left(Gm/\Delta E\right)^2$
and is independent of the relative velocity. Therefore, the importance of interactions of highly eccentric stars at distances much smaller than their semimajor axis, $r \ll a$, is proportional to $\rho(r)r$. Since the density of the cusp is steeper than $\rho \propto r^{-1}$, highly eccentric stars are mostly scattered close to their pericenter. If the velocity distribution is uniform, or, alternatively, if the distribution function is independent of the angular momentum, the fraction of stars that would have periapse distance $r \ll a$ is $r/a$. Therefore, these rare stars will dominate the global energy flux only if the cusp is steeper than $\rho \propto r^{-2}$.
The same conclusion can be achieved with the orbit-averaged Fokker-Planck equation (Eugene Vasiliev, private communication). The standard profile of \citet{baw76} has $1<\alpha=7/4<2$ and therefore, while the evolution of the few eccentric stars would be dominated by their interactions at small radii, their contribution to the total energy flux is small. However, we find that, with a source term, the slope external to the source is $\alpha=9/4>2$. Apparently, these stars, even though rare, dominate the overall energy flux and our expression for the energy flux (our Eq. (\ref{eqn:cuspnosour})) has to be modified. Yet, to globally change the profile, these stars have to be able to communicate with the majority of stars on the same seminajor axis, which do not have extremely eccentric orbits. This could not be done on a timescale shorter than the two body interaction time at that sememajor axis. Though a full treatment of this
effect is beyond the scope of this paper, we conjecture that the highly eccentric stars will evolve faster than the majority of stars and will not affect the distribution. The seemingly contradicting result that could be derived from the orbit average Fokker-Planck equation stems from the assumption that the distribution function will remain independent of the angular momentum.

\subsection{The case of binary disruption}

We now discuss how the process of binary disruption provides a source term within the cusp. A binary of total mass $M_{b}=m_1+m_2$ undergoes tidal breakup when passes inside the tidal radius
\begin{equation}
r_t=a_b\left(\frac{M_{BH}}{M_{b}}\right)^{1/3}\ ,
\end{equation}
where $a_b$ is the binary semimajor axis. One of the stars is expelled from the GC with velocities of hundreds km s$^{-1}$, while the other one remains bound to the SMBH \citep{hil88,yut03,sar10}.
The tidal radius becomes the pericenter of the captured star orbit but its semimajor axis is much larger, given by
\begin{equation}
r_{BH}\approx a_b\left(\frac{M_{BH}}{M_b}\right)^{2/3}
\label{eqn:rbin}
\end{equation}
If we assume that the binary semimajor axis are distributed according to \citep{duq91}
\begin{equation}
f(a_b)\propto \frac{1}{a_b}
\end{equation}
in the interval $[a_{min}$-$a_{max}]$, then, according to Eq. \ref{eqn:rbin}, the typical $r_{BH}$ of captured stars follows the same distribution scaled by a factor $Q^{2/3}=(M_{BH}/m)^{2/3}$
\begin{equation}
f(r_{BH})\propto\frac{Q^{2/3}}{r_{BH}}
\label{eqn:frbhbin}
\end{equation}
in the range $r_{min}=Q^{2/3} a_{min}$ and $r_{max}=Q^{2/3} a_{max}$. For the GC, \citet{yut03} showed that for binary tidal disruption by a SMBH the typical rate is $\approx 10^{-5}$-$10^{-4}$ yr$^{-1}$. In fact, up to a logarithmic factor, this is simply the inverse dynamical time at the radius of influence of the SMBH. Such theoretical prediction is also consistent with HVSs data \citep{brw15}. For what concerns the choice of the extremes of the semimajor axis interval, while the minimum value of $a_{min}$ is motivated by the minimum separation that two stars in a binary can have in order not to make them merge, the maximum value of $a_{max}$ deals with dynamical arguments. For solar mass stars, the minimum is $a_{min}=0.01$ AU. We introduce the adimensionless parameter
\begin{equation}
\beta=\frac{\epsilon}{m\sigma^2}=\frac{Gm}{2a\sigma^2}\ ,
\label{eqn:mamax}
\end{equation}
where $m$ is the star mass, $\epsilon$ is the absolute value of the negative internal energy of the binary and $\sigma$ is the velocity dispersion of stars. Binaries with $\beta\ll 1$ are called soft binaries and will become even softer on average until they are disrupted by the background population \citep{heg75}. Binaries with $\beta\gg 1$ are referred to as hard binaries and tend to become even harder by interacting with background stars \citep{heg75}. $\beta=1$ computed at $r_h$ gives approximatively the threshold between soft and hard binaries. In the case of Milky Way, $\hat{a}=0.1$ AU. As a consequence, binaries with initial semimajor axis $a\gg \hat{a}$ are typically disrupted by the background stars of the cusp before reaching the GC, while if $a\gg \hat{a}$ they can undergo tidal disruption by the SMBH \citep{hop09}.

Disruption of binaries therefore serve as a source term, with $\dot N_{inj}=\eta/P_h$, where $P_h=P(r_h)$ is the period at the influence radius. Instead of all being deposited at the same radius, these stars are distributed between $r_{min}$ and $r_{max}$. We estimate $\eta$ as a product of two factors:
\begin{equation}
\eta=\eta_b \omega .
\end{equation}
The first, $\eta_b$, is the fraction of binaries at the influence radius and the second, $\omega$, is the effective fraction of injected stars that circularize and therefore take part in the picture described above. $\eta_b$ is somewhat uncertain and probably observationally biased. \citet{pfu14} constrained the spectroscopic binary fraction of massive stars (OB and WR stars) in the GC to $0.30^{+0.34}_{-0.21}$, similar to the binary fraction in comparable young clusters. \citet{tok14a,tok14b} used a volume-limited sample of 4847 unevolved (or moderately evolved) stars within $\approx 67$ pc of the Sun with masses from $0.9\ \mathrm{M}_{\odot}$ to $1.5\ \mathrm{M}_{\odot}$ and found that $\approx 50$\% of stars have a stellar companion. \citet{hop09} found that the binary fraction at $r_h$ is $\approx 0.1$ in the GC by solving the Fokker-Planck equation for binary stars that interact with a static background of single stars. $\omega$ can be estimated in the following way. The captured star of the dissolved binary is on a highly eccentric orbit
\begin{equation}
e\approx 1-Q^{-1/3}\approx 0.99\ .
\end{equation}
As a consequence, the star can be rapidly tidally disrupted by the SMBH, if it enters the loss-cone, due to its initial highly-eccentric orbit. Moreover, the change in angular momentum that has to occur to disrupt the captured stars is much smaller than that to circularize its orbit. On the other hand, the change in angular momentum needed in order to circularize its orbit could be in an arbitrary direction, while the change in angular momentum needed to disrupt the binary has to be in a definite direction. To understand the relative fraction of these two channels, we can treat the evolution as a Brownian process governed by a continuous diffusion equation. Suppose stars are injected into angular momentum $J_0=(GM_{BH}r_{BH}(1-e^2))^{1/2}$, and from there each star can either diffuse into the lowest allowed angular momentum $J_{LC}$ (loss-cone) or go up towards that of a circular orbit $J_C$. We then solve the diffusion equation with null boundary condition at both size $f(J_{LC})=f(J_C)=0$, and get  (see similar considerations in \citet{wei17})
\begin{equation}
f(j)=f_{j_0}
\begin{cases}
ln (j/j_{LC}) \over \ln(J_0/J_{LC}) & {\rm for\,} j<J_0 \cr
\cr
ln (j/j_{C})  \over \ln(J_0/J_C) & {\rm for\,} j>J_0 
\end{cases}
\end{equation}
The constant $f_{j_0}$ is set by the injection rate, but it is of no interest to us. The ratio of fluxes upwards to circular orbits compared to that downward to tidal disruption is therefore
\begin{equation}
\omega=\frac{\Gamma_{circ}}{\Gamma_{disrupt}}=\frac{\ln(J_0/J_{LC})}{\ln(J_C/J_0)}=\frac{\ln(a_b/R_*)}{\ln(1/(1-e^2))}\approx 0.35
\end{equation}
Here, $R_*$ is the stellar radius. Hence, the captured star of most disrupted binaries is tidally disrupted, but still a significant fraction, $35\%$, are circularized. After circularization, they continue to change their semimajor axis by two body interactions. 

Putting all these consideration together (with $\omega \approx 0.35$), we obtain that the effective injection rate into tight circular orbits due to breakup of binaries is $\eta \approx 0.18$ for solar mass stars \citep{tok14a,tok14b}, $\eta \approx 0.10$ for massive stars \citep{pfu14}. In the case of \citet{hop09} model for the GC, $\eta \approx 0.035$.

\subsection{Cusp with binaries}

As a consequence of the presence of source term from binary disruption, the cusp may develop a steeper slope $\alpha=9/4$ out of the minimum injection radius $r_{min}$. The extent of such a region depends on $\eta$. We can use equation (\ref{eqn:cuspnosour}) and (\ref{eqn:cuspsour}) with $\dot N=\eta/P_h$ to obtain the cusp profile
\begin{equation}
N(r)={M \over m} \times 
\begin{cases}
\left(\eta r_h \over r_{min} \right)^{1/2} \left(r \over r_h\right)^{5/4} &\ {\rm for\,}\ r<r_{min} \\
\\
\eta^{1/2} \left(r \over r_h \right)^{3/4}  &\ {\rm for\,}\ r_{min}<r<\eta r_h \cr
\\
\left(r \over r_h \right)^{5/4} &\ {\rm for\,}\ \eta r_h<r
\end{cases}
\label{eqn:allcases}
\end{equation}

We can calculate the ratio between $N(r)$ as due to binary injection (from Eq. \ref{eqn:cuspsour}) and the standard $N_{BW}=N_h(r/r_h)^{5/4}$ \citep{baw76}, where $N_h=M/m$ is the number of particles at the influence radius, in the central region
\begin{equation}
\frac{N(r)}{N_{BW}(r)}=\eta^{1/2}\left(\frac{r}{r_h}\right)^{-1/2}\ .
\label{eqn:extent}
\end{equation}
As discussed, the cusp has slope $\alpha=7/4$ in the region $r<r_{min}$,
\begin{equation}
N(r)=N(r_{min})\left(\frac{r}{r_{min}}\right)^{5/4}\ ,
\end{equation}
with normalization given by
\begin{equation}
N(r_{min})=\eta^{1/2}\left(\frac{r_{min}}{r_h}\right)^{-1/2}N_{BW}(r_{min})\ .
\label{eqn:nrmin}
\end{equation}
The maximum extent of the region with a steeper cusp profile can be found by requiring that Eq. \ref{eqn:extent} is equal to unity
\begin{equation}
R_{max}=\eta\ r_h\ .
\end{equation}
Outside of $R_{max}$, the cusp turns back to the standard \citet{baw76} shape, while in the region $r_{min}<r<R_{max}$
\begin{equation}
N(r)=N(R_{max})\left(\frac{r}{R_{max}}\right)^{3/4}\ .
\end{equation}

Equation \ref{eqn:nrmin} allows to calculate also the minimum value of $\eta$ to have deviations in the cusp profile. By requiring that $N(r_{min})=N_{BW}(r_{min})$
\begin{equation}
\hat{\eta}=\frac{r_{min}}{r_h}\ .
\label{eqn:etamin}
\end{equation}
For $\eta>\hat{\eta}$, the cusp profile is modified by the injection of stars.

\section{Computational algorithm}

The most straightforward method to investigate the long-term effects of stars injection would be to integrate the system by direct $N$-body simulations \citep{bau04a,bau04b,tre07}. The main limitation of such an approach is the eccesive computational time forcing an unrealistic simulation limited to a small number of particles. The last attempt of pushing forward the limit of direct $N$-body simulations is by \citet{bau17}, but that simulation is still limited to $50$k stars.

In what follows, we describe the computational method we use to perform long-term evolution of the stellar cusp around the Milky Way's SMBH. Our scheme is conceptually similar to the method presented by \citet{hen71}. Our main aim is to investigate if the long-term injection of stars can modify the otherwise standard stellar cusp of \citet{baw76}. There are three main features of our simulation
\begin{itemize}
\item It follows the evolution of the semimajor axis of each star, therefore conserving the number of stars;
\item The 2-body scattering take energy from one star and give it to another, therefore conserving energy;
\item The rate of scattering changes with semimajor axis or energy, and with the number of particles that have similar energy.
\end{itemize}
We focus our attention on the region inside the sphere of influence, where the stellar dynamics is strongly influenced by the extreme field of the SMBH. In all our calculations, we assume $M_{BH}=4\times 10^6\ \mathrm{M}_{\odot}$ \citep{gil17} and a single-mass population of stars of mass $m_*=1\ \mathrm{M}_{\odot}$. The region of interest span a wide range of energy and distances with respect to the SMBH. While the outermost radius is taken to be the SMBH influence radius
\begin{equation}
r_h=\frac{GM_{BH}}{\sigma^2}\approx 2\ \mathrm{pc}\ ,
\end{equation}
where $\sigma$ is the velocity dispersion external to the radius of influence, the innermost radius is taken to be equal to the tidal disruption radius of $1\ \mathrm{M}_{\odot}$ star \citep{sto13}
\begin{equation}
r_{in}= R_*\left(\frac{M_{BH}}{m}\right)^{1/3}\approx 1\ \mathrm{AU}\ .
\end{equation}

We now introduce the system of units we use in our code. In such a system, $G=1$, $M_{BH}=1$ and $r_h=1$, and $r_h$ and $r_{in}$ correspond to the dimensionless energies $E_{out}=E_{h}=GM_{BH}/(2r_{h})=0.5$ and $E_{in}=GM_{BH}/(2r_{in})=10^5$, respectively. Each particle has a semimajor axis and a corresponding energy. The energy of the stars is a continuous variable, but we devide the stars into energy bins. Each energy bin is an interval between $E_{i-1}$ and $E_i$, where we use $E_i/E_{i-1}=2$. At the beginning of the simulation, we set each our stars with some initial energy and then count how many of them are in each energy bin. The bins are used only in order to estimate the local $2$-body scattering rate
\begin{equation}
\Gamma_i=\frac{\zeta}{P_h}\left(\frac{m}{M_{BH}}\right)^2 N_i \ln\left(\frac{M_{BH}}{m}\right) \left(\frac{E_i+E_{i-1}}{2 E_h}\right)^{3/2}\ ,
\end{equation}
where $N_i$ is the number of stars in the energy interval $[E_{i-1}$-$E_i]$. In the above equation
\begin{equation}
P_h=2\pi\frac{r_h^{3/2}}{(GM_{BH})^{1/2}}=2\pi
\end{equation}
is the period at the influence radius, while $\zeta=\gamma/\xi^2$. The factor $\xi$ is the exchanged energy fraction (see equations (\ref{energychange}) below) and
\begin{equation}
\gamma=\frac{E_i+E_{i-1}}{2(E_i-E_{i-1})}
\end{equation}
is a factor of order of unity that takes into account the bin size. In our case, $\gamma=1.5$. The factor $\xi^{-2}$ takes into account the small scatterings are more frequent than the strong ones. We take a time-step of
\begin{equation}
\Delta t=\frac{\psi}{\max\limits_{i} \Gamma_i}\ .
\end{equation}
so that no where in our energy grid there are more scatterings than particles. The number of $2$-body scattering events in each bin is taken to be
\begin{equation}
N_{ev,i}=N_i \Gamma_i \Delta t\ .
\end{equation}
In our simulations, we set $\psi=0.5$. The value of $\psi$ limits the maximum number of scattered particles to $\psi N_i$ in the bin in which $\Gamma_i=\max \Gamma_i$. For each scattering event, two random particles, with energy $E_1$ and $E_2$ respectively, are chosen among the $N_i$ particles of the bin. Their energies are updated according to 
\begin{eqnarray}
\label{energychange}
E_{1,f}&=&E_{1,i}+\chi\xi(E_{1,i}+E_{2,i})\\
E_{2,f}&=&E_{2,i}-\chi\xi(E_{1,i}+E_{2,i})\ ,
\end{eqnarray}
where $\chi$ is a random variable between $0$ and $1$ and $\xi$ is the maximal fraction of exchanged energy. We set $\xi=0.1$. After the $N_{ev,i}$ events in each bin, if the particle energy $E>E_{in}$ the particle is removed from the calculation (accretion onto the SMBH), while if $E<E_h$ (escaping the cusp), the particle is replaced by a particle with a random energy in the last bin. This replacement mimics the isothermal sphere which surrounds the radius of influence, which contains stars that could become bound \citep{hop06,hoa06}. Stars may also be disrupted by the SMBH if their angular momentum is smaller than $J_{LC}$, and diffusion in the angular momentum space is more efficient than in the energy space \citep{ale17}. A complete treatment of the loss-cone problem would require a two-dimensional approach in both energy and angular momentum space. An approximate solution can be obtained by considering only the energy space and by adding a sink term \citep{lig77,sha78}. \citet{hoa06} showed that the presence of such an effective loss-cone sink term does not change the star distribution. In order to mimic the loss-cone sink effect, we remove in each bin the fraction of stars with angular momentum larger than $J_{LC}$. Such fraction is simply $J_{LC}/J_C$ if we assume a constant distribution of angular momenta. We also find the loss-cone sink effect does not change the overall star distribution.

As discussed in the previous section, we parameterize the rate of disrupted binaries with the dimensionless parameter $\eta$. After updating the energies of all the particles according to the scheme described above, we generate
\begin{equation}
N_b=\eta\ \Delta t /P_h
\label{eqn:binj}
\end{equation}
injected stars as a consequence of the binary tidal disruption in each timestep. The energies of such stars are computed after sampling their position according to Eq. \ref{eqn:frbhbin} and are added to the pre-existing population. If $\eta=0$, there is no injection of stars from disrupted binaries. Note that $\eta$ also parametrizes the fraction of injected stars that are not tidally disrupted. $\eta=1$ corresponds to the case where all stars at the radius of influence are binaries and all the captured stars from a dissolved binary are circularized, and is therefore an upper limit to the rate of stellar injection.

\section{Simulation results}

\begin{table}
\caption{Models: name, binaries, minimum injection radius ($r_{min}$), maximum injection radius ($a_{max}$), binary fraction ($\eta$).}
\centering
\begin{tabular}{lcccc}
\hline
\hline
Name & Binaries & $r_{min}$ (AU) & $r_{max}$ (AU) & $\eta$\\
\hline
Model 0	    & no	& -		 		& -			& $0$ \\
Model 1		& yes	& $250$ 		& $2500$ 	& $0.01$-$1$ \\
Model 1-1	& yes	& $250$		 	& $5000$	& $0.5$ \\
Model 2		& yes	& $250$-$1250$ 	& $2500$	& $0.5$ \\
\hline
\end{tabular}
\label{tab1}
\end{table}

We consider three different models, as summarized in Table \ref{tab1}. In Model 0, we consider no injection of stars ($\eta=0$). This should result in the standard \citet{baw76} cusp, i.e. $N(r)=N_h\ r^{5/4}$. We use that as a check for our numerical procedure. In Models 1 and 2, we inject stars as a consequence of binary tidal disruptions, with different rates and different minimal injection radius corresponding to a minimal binary separation $a_{min}$. In Model 1, we take the minimum initial semimajor axis $a_{min}=0.01$ AU for solar mass stars, while $a_{min}=0.05$ AU in Model 2. In all the simulations with $\eta>0$, we keep $a_{max}=0.1$ AU fixed, as we found that it has little influence on the outcome. However, we run an additional simulation with $a_{max}=0.5$ AU to check that the maximum injection radius has negligible effect on the outcomes (Model 1-1).

\begin{figure*} 
\centering
\begin{minipage}{18cm}
\includegraphics[scale=0.70]{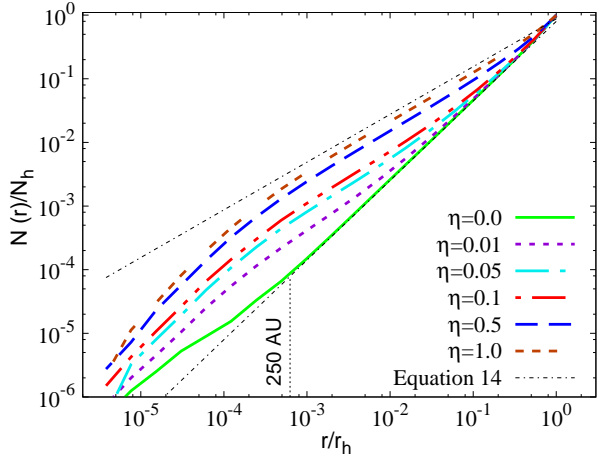}
\includegraphics[scale=0.70]{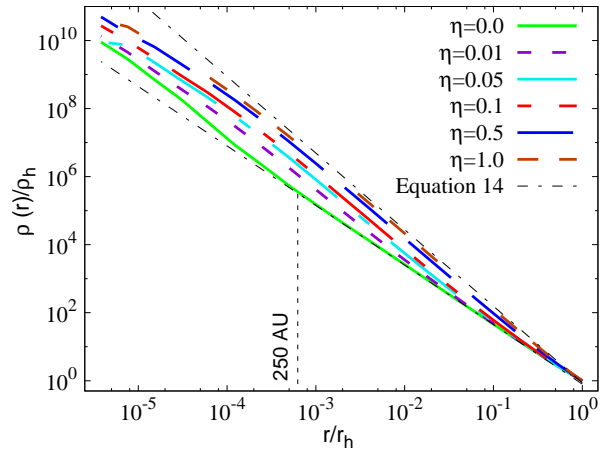}
\caption{Cusp profile for Model 0 ($\eta=0$) and Model 1. The curves $N(r)/N_h=(r/r_h)^{5/4}$ (left) and $\rho(r)/\rho_h=(r/r_h)^{-7/4}$ (right), i.e. the standard \citet{baw76} profile, and Eq. \ref{eqn:allcases} and the corresponding density for $\eta=1$ are shown as reference. The slope of the cusp depends on the injection rate. We also show the minimum injection radius $r_{min}=250$ AU, corresponding to $a_{min}=0.01$ AU.}
\label{fig:cusp1}
\end{minipage}
\end{figure*}

Figure \ref{fig:cusp1} shows the profile $N(r)/N_h$ (left panel) and $\rho(r)/\rho_h$ (right panel) of the simulations for Model 0 and Model 1, along with the theoretical prediction curves $N(r)\propto r^{5/4}$ and $\propto r^{3/4}$ (left panel) and $\rho(r)\propto r^{-7/4}$ and $\propto r^{-9/4}$ (right panel), according to Eq. \ref{eqn:allcases}. The total time of the simulation may depend on the choice of the initial conditions. Initial conditions very far from the equilibrium take more time to converge to a steady-state solution. We start with all the stars at low energies as in \citet{hop09} (see also \citet{mad11}). All the simulations are evolved until a timescale of the order of $\approx T_h$, where $T_h$ is the $2$-body relaxation timescale at the influence radius. In particular, we initialize all the stars in the bin containing $r_h/2$. Initializing all the particles with high energies may lead to a longer timescale for the simulations. Actually, since $T_{h,i}\propto N^
{-1}$ in each bin, only few particles are available in the last bins to be relaxed until some of them are supplied from smaller radii. Moreover, $T_{h,i}\propto r^
{3/2}$. As a consequence, starting with all the particles at high energies makes the total time to form a cusp within $r_h$ longer because the local relaxation time far from the SMBH, near the influence radius, is longer. Stars at any other distance are effectively relaxed within $\approx T_h$. For $\eta=0$ (no injection of binaries), our results recover the standard \citet{baw76} cusp profile as expected, where conservations of the energy leads to the $\alpha=7/4$ slope (see solid green line in Fig. \ref{fig:cusp1}).

\begin{figure}
\centering
\includegraphics[scale=0.70]{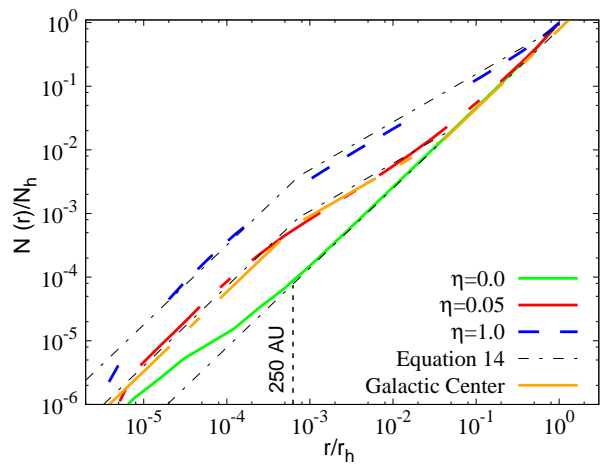}
\caption{Analytical curves for the distribution of stars from Eq. \ref{eqn:allcases} and simulation results in the cases of the standard \citet{baw76} cusp and $\eta=0.05$ and $\eta=1$. We also show the predicted distribution for the GC ($\eta=0.035$).}
\label{fig:cusp2}
\end{figure}

We also illustrates in Fig. \ref{fig:cusp1} the results of simulations if stars are injected as a consequence of tidal binary disruption for different values of $\eta$. As discussed, the cusp has the same slope of the standard one in the region $r<r_{min}$, since a constant energy flow still dictates $N(r) \propto r^{5/4}$ (see all dashed lines in Fig. \ref{fig:cusp1}). However, the overall normalization changes as a consequence of stars injection according to Eq. \ref{eqn:allcases}. As defined in Eq. \ref{eqn:binj}, $\eta$ corresponds to the binary fraction times the fraction of captured stars that are circularized. In the case of the GC, the typical binary tidal disruption rate is $\approx \eta_b/P_h$ \citep{yut03,sar10}, which may be increased by massive perturbers \citep{pha07}. According to $\eta$, the cusp profile becomes steeper ($\alpha=9/4$) in the region $r_{min}<r<R_{max}$, while turns back to the standard \citet{baw76} shape beyond $R_{max}$. For what concerns the evolution in time, initially, the cusp starts deviating from the standard $\alpha=7/4$ cusp in the injection point, and this will slowly spread outward according to $\eta$. Moreover, we note that the extent of the steeper cusp will not change in time as long as binary are injected and disrupted by the SMBH.

In Fig. \ref{fig:cusp2}, we show the analytical predictions for the distribution of stars obtained from Eq. \ref{eqn:allcases} along with the results from our simulations in the cases $\eta=0.05$ and $\eta=1$. We also illustrate the predicted distribution for the GC ($\eta=0.035$). In this models $r_{min}=250$ AU, hence the normalization $N(r_{min})=40 \eta^{1/2} N_{BW}(r_{min})$. The simulated profiles are consistent with the analytical predictions. In the case $\eta=1$, $R_{max}=r_h$ and the region with the steeper slope $\alpha=9/4$ extends up to the influence radius. In the case $\eta=0.5$, the cusp is steeper up to $R_{max}=0.05 r_h$, while returns to the \citet{baw76} beyond it.

\begin{figure}
\centering
\includegraphics[scale=0.70]{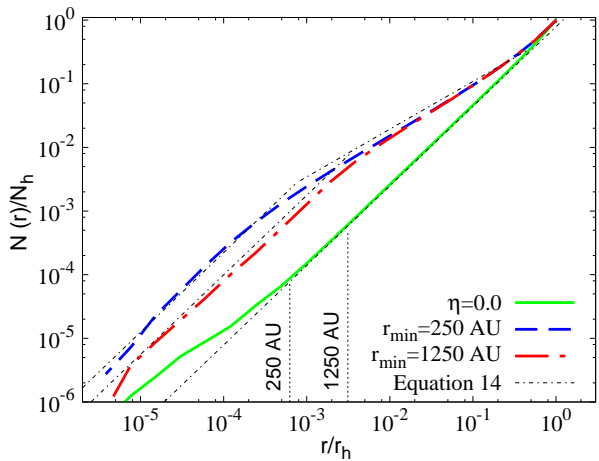}
\caption{Cusp profile for Model 0 ($\eta=0$) and Model 2. The theoretical curves $N(r)$ from Eq. \ref{eqn:allcases} are shown as reference. The break in the distribution depends on the minimum injection radius. We show also the minimum injection radii $r_{min}=250$ AU and $r_{min}=1250$ AU, corresponding to $a_{min}=0.01$ AU and $0.05$ AU, respectively.}
\label{fig:cusp3}
\end{figure}

Figure \ref{fig:cusp3} shows the effect of changing the minimum injection radius (Model 2). We run two simulations with $r_{min}=250$ AU and $r_{min}=1250$ AU, respectively, when $\eta=0.5$. As discussed previously, in the inner region ($r<r_{min}$), the inward flux is identical to the \citet{baw76} case, and $N(r) \propto r^{5/4}$. The slope of the distribution changes outside of $r_{min}$. As also showed in Fig. \ref{fig:cusp3}, $\eta=0.5$ leads to an outer region with slope $\alpha=3/4$ outside of $\approx 250$ AU and $\approx 1250$ AU, respectively. Moreover, different minimum injection radii give different normalizations according to Eq.~\ref{eqn:nrmin}. The theoretical predictions are consistent with our numerical results.

We also run an additional simulation (Model 1-1) with $a_{max}=0.5$. and checked that our results do not depend on the choice of $a_{max}$.

\section{The Stellar Nucleus of The Milky Way}

Our models show that binary tidal disruption can change the slope of the cusp external to the minimum injection radius and lead to a steeper slope $\alpha=9/4$ when the injection rate is high enough. Stellar cusps in galactic nuclei have been quite elusive observationally, mainly because SMBH radii of influence are barely resolved because of their large distances. As a consequence, a few bright stars can prevent the determination of the real cluster structure, and most faint stars may not be detected due to source crowding because of the large stellar densities. Moreover, the high interstellar extinction allow to observe stars only in the near-infrared with good sensitivity \citep{sch14,sga17}.

Clearly, the best data to describe the structure and dynamics of stars in a galactic nucleus are available only for our own GC thanks to its relative vicinity.
Data have usually been limited to the red clump stars and bright giants, which represents only a small fraction of the population of stars in the GC \citep{sch14}.  Recently, \citet{gal17} and \citet{sch17} have used observational data taken with adaptive optics with the NACO instrument at the ESO VLT. They studied the old population of $1$-$2\ \mathrm{M}_{\odot}$ stars and found that the 3D stellar distribution is well fitted by a power-law profile with slope $\approx 1.2$ within $\approx 3$ pc. Such a slope is even shallower than the standard slope of \citet{baw76} cusp.
 \citet{sch17} also found that giant stars have a cored profile within a $\approx 0.3$ pc of the SMBH, with a cuspy distribution at larger distances, in agreement with previous observations \citep{doo09,btk10}. Such cored profile may be due to stellar collisions \citep{dal09}, the presence of a fragmented gaseous disk \citep{ama14} or subsequent epochs of star formation \citep{aha15}.

\citet{hop09} showed that the binary fraction at $r_h$ is $\approx 0.1$ implying $\eta\approx 0.035$. As shown in Fig. \ref{fig:cusp2}, such $\eta$ is enough to modify the cusp within $r<\eta r_h=\approx 0.07$ pc. Assuming as minimum injection radius $r_{min}=250$ AU, the binary-modified profile would be $\approx 8$ times denser than the standard cusp (see Eq. \ref{eqn:extent}) at $r<r_{min}$. On the other hand, \citet{gal17} and \citet{sch17} have recently constrained the cusp slope $\approx 1.2$ within $\approx 3$ pc, even shallower than the standard \citet{baw76} cusp. Several factors may be responsible for the discrepancy between our model and the observations. First, on the observational side, distinguishing between a single power-law profile and a broken power-law within $\approx 0.07$ pc is not straightforward. Second, on the theoretical part, we do not account for the stellar mass distribution, which tends to flatten the cusp \citet{bau17}. Third, we have not taken into account the finite lifetime of massive stars, while most observed ones (S-stars) are massive \citep{gil17}. 

We note that in our model we do not take into account neither the coherent torques between slowly precessing orbits, i.e. resonant relaxation, nor energy loss as a consequence of gravitational waves \citep{ant12,ant13}. Such mechanisms are expected to be important in the innermost region near the SMBH, $r\lesssim 2000$ AU \citep{hoa06}, as in the case of the S-stars whose orbits may have been shaped by the resonant relaxation process \citep{pgk09,pgu10,ale17,gil17}. We showed that the standard profile is modified out of $r_{min}$ if $\eta$ is sufficiently high. As a consequence, the region $2000\ \mathrm{AU}\lesssim r\le \eta r_h$ should still present a modified cusp as due to binary injection.

We have not considered a mass function both for the stars in the cusp and the injected stars, but only a single-mass population of $1\ \mathrm{M}_{\odot}$ stars. As shown by \citet{ale09} and \citet{bau17}, different stellar types have different slopes, where stellar black holes have the largest value $\alpha_{BH}\approx 1.5$-$2$. If the slope is even steeper than predicted in the present work for the stellar black hole population then a larger number of stellar black holes in the cusp can form binaries and merge via gravitational wave emission as a consequence of the Kozai-Lidov effect, suggesting a higher rate of gravitational waves events \citep{ant12,hoa17}. The picture is even complicated by the possible presence of intermediate mass black holes and other remnants brought by inspiralling star clusters \citep{fag17,fgk17}.

In the near future, the spacetime of the Milky Way's SMBH may be probed thanks to next generation instruments such as \textit{GRAVITY}. The hope is to provie observational support to the no-hair theorem by monitoring the relativistic effects in the orbits of stars orbiting the SMBH inside $\approx 2000\ GM_{BH}/c^2\approx 100$ AU \citep{wil08,psa16}. \citet{mer10} studied the conditions under which the relativistic effects (frame-dragging and quadrupole precessions) of the Milky Way's SMBH can be measured by \textit{GRAVITY}  and found that detection of frame-dragging precession can be achieved by monitoring stars between $\approx 40$ AU and $\approx 100$ AU for few years. Quadrupole-induced precession can be observed only under fine-tuned conditions for $r\lesssim 40$ AU. Currently, the lowest well measured approach to is of S2 $\approx 100$ AU \citep{gil17}. If we assume $\eta\approx 0.035$ as discussed before, we would get that our model predicts $\approx 1200$ stars within $\approx 120$ AU, $\approx 8$ times larger than the standard cusp. This would be an encouraging prediction for GRAVITY, impling a larger number of stars useful to measure relativistic precession. Yet, a firm prediction regarding our own GC will have to take into account the stellar mass function and resonant relaxation.

\section{Conclusions}

The distribution of stars around a SMBH has been under intense scrutiny since the \citet{baw76} canonical cusp model. In this model, the energy conservation governs the overall slope of cusp leading to the standard value $\alpha=7/4$. Binary tidal disruption by SMBHs has attracted attentions since the \citet{hil88} paper, and the discovery of hypervelocity stars \citep{brw15}. In this paper, we study the long-term effect of the process of binary breakup by adding a source term in the standard cusp picture. By means of theoretical arguments and numerical simulations, we show that such a source term changes the slope of the cusp external to the source position leading to a steeper slope $\alpha=9/4$ if the rate is high enough.

If we assume that the binary fraction in the GC is $0.1$ \citep{hop09}, we would get $\eta\approx 0.035$. As shown in Fig. \ref{fig:cusp1}, such $\eta$ is enough to modify the cusp within $\approx 0.07$ pc. Assuming as minimum injection radius of $r_{min}=250$ AU, the binary-modified profile would be $\approx 8$ times denser than the standard cusp (see Eq. \ref{eqn:extent}) at $r<r_{min}$. 

\section{Acknowledgements}
This research was partially supported by an ISF and an iCore grant. We acknowledge Tal Alexander, Bence Kocsis, Scott Tremaine, Eugene Vasiliev and Ben Bar-Or for useful discussions, and anonymous referee for constructive comments.



\begin{thebibliography}{99}
\bibitem[Aharon \& Perets(2015)]{aha15} Aharon, D., \& Perets, H. B. 2015, ApJ, 799, 185
\bibitem[Alexander(2017)]{ale17} Alexander, T. 2017, Ann. Rev. Astron. Astrophys., 55
\bibitem[Alexander \& Hopman(2009)]{ale09} Alexander, T., \& Hopman, C. 2009, ApJ, 697, 1861
\bibitem[Alexander \& Livio(2001)]{ale01} Alexander, T. \& Livio, M. 2001, ApJ Lett., 560, L143
\bibitem[Amaro-Seoane \& Chen(2014)]{ama14} Amaro-Seoane, P. \& Chen, X. 2014, ApJ, 781, L18
\bibitem[Antonini \& Merritt(2013)]{ant13} Antonini, F., \& Merritt, D. 2013, ApJ Lett., 763, L10
\bibitem[Antonini \& Perets(2012)]{ant12} Antonini, F., \& Perets, H. B. 2012, ApJ, 757, 27
\bibitem[Bahcall \& Wolf(1976)]{baw76} Bahcall, J. N., \& Wolf, R. A. 1976, ApJ, 209, 214
\bibitem[Bahcall \& Wolf(1977)]{baw77} Bahcall, J. N., \& Wolf, R. A. 1977, ApJ, 216, 883
\bibitem[Bar-Or \& Alexander(2014)]{bar14} Bar-Or, B., Alexander, T., 2014, Class. Quant. Grav., 31, 244003
\bibitem[Bar-Or \& Alexander(2016)]{bar16} Bar-Or, B., Alexander, T., 2016, ApJ, 820, 129
\bibitem[Bar-Or et al.(2013)]{bar13} Bar-Or B., Kupi G, \& Alexander T. 2013, ApJ 764, 52
\bibitem[Bartko al.(2010)]{btk10} Bartko, H. et al. 2010, ApJ, 708, 834
\bibitem[Baumgardt et al.(2004a)]{bau04a} Baumgardt, H., Makino, J., \& Ebisuzaki, T. 2004a, ApJ, 613, 1133
\bibitem[Baumgardt et al.(2004b)]{bau04b} Baumgardt, H., Makino, J., \& Ebisuzaki, T. 2004b, ApJ, 613, 1143
\bibitem[Baumgardt et al.(2017)]{bau17} Baumgardt, H., Amaro-Seoane, P., \& Sch\"{o}del R., 2017, preprint, arXiv:1701.03818
\bibitem[Binney \& Tremaine(2008)]{bin08} Binney, J., \& Tremaine, S. 2008, Galactic Dynamics (Princeton, NJ: Princeton Univ. Press)
\bibitem[Brown(2015)]{brw15} Brown W. R. 2015, ARA\& A, 53, 15
\bibitem[Capuzzo-Dolcetta \& Fragione(2015)]{cap15} Capuzzo-Dolcetta, R., \& Fragione, G. 2015, MNRAS, 454, 2677
\bibitem[Dale et al.(2009)]{dal09} Dale, J. E., Davies, M. B., Church, R. P., \& Freitag, M. 2009, MNRAS, 393, 1016
\bibitem[Do et al.(2009)]{doo09} Do, T. et al. 2009, ApJ, 703, 1323
\bibitem[Duquennoy \& Mayor(1991)]{duq91} Duquennoy, A., \& Mayor, M. 1991, A\&A, 248, 485
\bibitem[Fragione, Antonini \& Gnedin(2017)]{fag17} Fragione, G., Antonini, F., \& Gnedin, O. Y. 2017, preprint, arXiv:1709.03534
\bibitem[Fragione \& Capuzzo-Dolcetta(2016)]{fra16} Fragione, G., \& Capuzzo-Dolcetta, R. 2016, MNRAS, 458, 2596
\bibitem[Fragione, Ginsburg \& Kocsis(2017)]{fgk17} Fragione, G., Ginsburg, I., \& Kocsis, B. 2017, preprint, arXiv:1711.00483
\bibitem[Fragione, Capuzzo-Dolcetta \& Kroupa (2017)]{fck17} Fragione, G., Capuzzo-Dolcetta, R., \& Kroupa, P. 2017, MNRAS, 467, 451
\bibitem[Fragione \& Ginsburg(2017)]{frg17} Fragione, G., \& Ginsburg, I. 2017, 466, 1805
\bibitem[Fragione \& Loeb(2017)]{frl17} Fragione G., \& Loeb A., 2017, New Astronomy, 55, 32
\bibitem[Gallego-Cano et al.(2017)]{gal17} Gallego-Cano, E. et al. 2017, preprint, arXiv:1701.03816
\bibitem[Gillessen et al.(2017)]{gil17} Gillessen, S., Plewa, P. M., Eisenhauer, F., Sari, R. et al. 2017, ApJ, 837, 30
\bibitem[Ginsburg et al.(2012)]{gin12} Ginsburg, I., Loeb, A.,\& Wegner, G. A. 2012, MNRAS ,423.1, 948
\bibitem[Gnedin et al.(2005)]{gne05} Gnedin, O. Y., Gould, A., Miralda-Escud\'{e}, J.,\& Zentner, A. R. 2005, ApJ, 634, 344
\bibitem[Gould \& Quillen(2003)]{gou03} Gould A., Quillen A. C. 2003, ApJ, 592, 935
\bibitem[Heggie(1975)]{heg75} Heggie, D. C. 1975, MNRAS, 173, 729
\bibitem[H\'{e}non(1971)]{hen71} H\'{e}non, M. 1971, ApSS, 14, 151
\bibitem[Hills(1988)]{hil88} Hills, J. G. 1988, Nature, 331, 687
\bibitem[Hoang et al.(2017)]{hoa17} Hoang, B. M. et al. 2017, preprint arXiv:1706.09896
\bibitem[Hopman(2009)]{hop09} Hopman, C. 2009, ApJ, 700, 1933
\bibitem[Hopman \& Alexander(2005)]{hop05} Hopman, C., \& Alexander, T. 2006, ApJ, 629, 362
\bibitem[Hopman \& Alexander(2006a)]{hop06} Hopman, C., \& Alexander, T. 2006, ApJ, 645, 1152
\bibitem[Hopman \& Alexander(2006b)]{hoa06} Hopman, C., \& Alexander, T. 2006, ApJ Lett., 645, L133
\bibitem[Hopman et al.(2004)]{hop04} Hopman, C., Portegies Zwart, S. F., \& Alexander, T. 2004, ApJ, 604, L101
\bibitem[Kobayashi et al.(2012)]{kob12} Kobayashi, S., Hainick, Y., Sari, R. \& Rossi, E. M. 2012, ApJ, 748, 105 
\bibitem[Kroupa(2001)]{kro01} Kroupa P., 2001, MNRAS, 322, 231
\bibitem[Lightman \& Shapiro(1977)]{lig77} Lightman A. P., \& Shapiro, S. L. 1977, ApJ, 211, 244
\bibitem[Madigan et al.(2011)]{mad11} Madigan, A. M., Hopman, C., \& Levin, Y., 2011, ApJ, 738, 99
\bibitem[Merritt et al.(2010)]{mer10} Merritt, D., Alexander, T., Mikkola, S., \& Will, C. M. 2010, Phys. Rev. D, 816, 062002
\bibitem[Merritt(2013)]{mer13} Merritt, D. 2013, Dynamics and evolution of galactic nuclei. Princeton University Press
\bibitem[Merritt \& Ferrarese(2001)]{mer01} Merritt, D., \& Ferrarese, L. 2001, ApJ, 547, 140
\bibitem[Peebles(1972)]{pee72} Peebles, P. J. E. 1972, ApJ,178, 371
\bibitem[Perets \& Gualandris(2010)]{pgu10} Perets, H. B., \& Gualandris, A. 2010, ApJ, 719, 220
\bibitem[Perets et al.(2009)]{pgk09} Perets, H. B., Gualandris, A., Kupi, G., Merritt, D., \& Alexander, T. 2009, ApJ, 702, 884
\bibitem[Perets et al.(2007)]{pha07} Perets, H. B., Hopman, C., \& Alexander, T. 2007, ApJ, 656, 709
\bibitem[Pfuhl et al.(2014)]{pfu14} Pfuhl, O. et al. 2014, ApJ, 782, 101
\bibitem[Pfuhl et al.(2011)]{pfu11} Pfuhl, O., Fritz, T. K., Zilka, M., et al. 2011, ApJ, 741, 108
\bibitem[Psaltis et al.(2016)]{psa16} Psaltis, D., Wex, N., \& Kramer, M. 2016, ApJ, 818, 121
\bibitem[Rossi et al.(2014)]{ros14} Rossi, E. M., Kobayashi, S., \& R. Sari 2014, ApJ, 795, 125 
\bibitem[Rossi et al.(2017)]{ros17} Rossi, E. M., et al. 2017, MNRAS, 467, 1844
\bibitem[Sari et al.(2010)]{sar10} Sari, R., Kobayashi, S.,\& Rossi, E. M. 2010, ApJ, 708, 605
\bibitem[Sari \& Goldreich (2006)]{sag06} Sari, R. \& Goldreich, P. 2016, ApJ, 642, 65
\bibitem[Sch\"{o}del et al.(2014)]{sch14} Sch\"{o}del, R., Feldmeier, A., Neumayer, N., Meyer, L., \& Yelda, S., 2014, Classical and Quantum Gravity, 31, 244007
\bibitem[Sch\"{o}del et al.(2017a)]{sch17} Sch\"{o}del, R. et al. 2017, preprint, arXiv:1701.03817
\bibitem[Sch\"{o}del et al.(2017b)]{sga17} Sch\"{o}del, R., Gallego-Cano, E., \& Amaro-Seoane, P. 2017, preprint, arXiv:1702.00219
\bibitem[Shapiro \& Lightman(1976)]{sha76} Shapiro S. L., \& Lightman A. P. 1976, Nature 262, 743
\bibitem[Shapiro \& Marchant(1978)]{sha78} Shapiro S. L., \& Marchant, A. B. 1978, ApJ 225, 603
\bibitem[Stone et al.(2013)]{sto13} Stone, N., Sari, R., \& Loeb, A. 2013, MNRAS, 435, 1809
\bibitem[Tokovinin(2014a)]{tok14a} Tokovinin A. 2014, AJ, 147, 86
\bibitem[Tokovinin(2014b)]{tok14b} Tokovinin A. 2014, AJ, 147, 87
\bibitem[Trenti et al.(2007)]{tre07} Trenti, M., Heggie, D. C., \& Hut, P. 2007, MNRAS, 374, 344
\bibitem[Weissbein \& Sari(2017)]{wei17} Weissbein, A., \& Sari, R. 2017, MNRAS, 468, 1760
\bibitem[Will(2008)]{wil08} Will, C. 2008, ApJ Lett., 674, L25
\bibitem[Yu \& Tremaine(2003)]{yut03} Yu, Q., \& Tremaine, S. 2003, ApJ, 599, 1129
\bibitem[Zubovas et al.(2013)]{zub13} Zubovas K., Wynn G. A., \& Gualandris A. 2013, ApJ, 771, 118
\end{thebibliography}
\end{document}